\documentclass[journal]{IEEEtran}

\usepackage{graphicx}
\usepackage{xcolor}
\usepackage{stfloats} 
\usepackage{cite}
\usepackage{amsmath}
\usepackage{url}

\begin{document}

\title{
%Fault-ride-through control of offshore AC islands with HVDC in-feeds

Fault-Ride-Through Coordination Strategy \\ for Offshore AC Islands with \\ Multi-Infeed HVDC Interconnections
}

\author{Eleni~Tsotsopoulou, Vasileios~Psaras, Dionysios~Moutevelis,~\IEEEmembership{Member,~IEEE},
\\
Oriol Gomis-Bellmunt,~\IEEEmembership{Fellow,~IEEE,} and Alexandros~Paspatis,~\IEEEmembership{Senior Member,~IEEE}% <-this % stops a space

\thanks{E. Ttsotsopoulou and V. Psaras are with WSP in the UK, UK (eleni.tsotsopoulou@wsp.com, vasileios.psaras@wsp.com). D. Moutevelis and O. Gomis-Bellmunt are with Universitat Politècnica de Catalunya, Spain (dionysios.moutevelis@upc.edu, 
oriol.gomis@upc.edu) A. Paspatis is with the Manchester Metropolitan University, UK (a.paspatis@mmu.ac.uk). 
} 
\thanks{
Corresponding Author: Alexandros Paspatis (a.paspatis@mmu.ac.uk).
}
}

% Journal name
\markboth{}%
{}

\maketitle

\begin{abstract}
Large-scale offshore Wind Farms (WFs) are considered key assets towards realizing a sustainable power system.
These systems are often configured as offshore AC islands and their integration largely depends on the High-Voltage-Direct-Current (HVDC) technology. 
This topology, while it enables cost-effective transmission over large offshore distances, may lead to operational challenges.
Specifically, the operation of offshore AC islands during faults and the grid code requirement fulfillment are identified as a major challenges for their large-scale deployment.
To address this pressing issue, a comprehensive coordination control strategy for the different participating converters in multi-infeed AC offshore islands  during Fault Ride Through (FRT) operation is presented in this work.
The proposed strategy introduces advanced control functions in the FRT schemes of both the HVDC and WF converters, such as zero active and reactive power injection during faults, as well as post-fault active power droop
%P/f and f(P) droop
control coordination to tackle power imbalances.
The proposed FRT coordination strategy is validated through both extensive simulations in PSCAD/EMTDC, as well as with Power Hardware-in-the-Loop (PHIL) experimental results, considering both AC and DC faults.
\end{abstract}


\IEEEpeerreviewmaketitle

\section{Introduction}

\IEEEPARstart{T}{he} European Union has set ambitious decarbonisation targets, aiming for net zero emissions by 2050 \cite{b1}. As electricity generation has historically been a major contributor to greenhouse gas emissions, Renewable Energy Sources (RES) have become central to achieving these goals. However, large scale RES integration remains constrained by technical, regulatory, social, and financial limitations~\cite{psaras2020non}. In many regions, RES hosting capacity is already saturated, and further deployment requires significant network reinforcement.

To overcome these limitations, transmission system interconnection, primarily through High Voltage Direct Current (HVDC) links based on Modular Multilevel Voltage Source Converters (MMC-VSC), has emerged as a key enabler of flexibility~\cite{van2016hvdc}. HVDC technology allows power to be exchanged across wide geographical areas, smoothing variability in generation and demand while relieving network bottlenecks. At the same time, spatial constraints and the high power density of modern renewable plants have accelerated the shift toward offshore wind generation. The combination of offshore Wind Farms (WF) and sub-sea HVDC interconnectors is therefore playing a pivotal role in Europe’s energy transition~\cite{korompili2016review}.

A major development in this direction is the emergence of offshore AC energy islands, i.e., large offshore hubs that collect wind power and interconnect multiple, possibly asynchronous, power systems via HVDC links~\cite{Lit_1}.
This concept is central to the UK’s Holistic Network Design (HND) and Beyond 2030 programmes, which aim to deliver the Government’s ambition of connecting more than 50~GW of offshore wind in the coming decade \cite{HND1, HND2, HND3}. The proposed network design includes multiple HVDC links arriving at the same AC energy island, forming the so-called multi-infeed AC islands in the North Sea. Similar initiatives are underway in neighbouring countries: Denmark is developing two major energy islands —an artificial island in the North Sea (initially 3~GW, expandable to 10~GW)~\cite{Lit_1}, and the Bornholm island hub in the Baltic Sea (3~GW by 2030) \cite{Case2}; while Belgium is constructing the Princess Elisabeth Energy Island, a 3.5~GW hybrid AC/DC hub with planned HVDC links to Denmark and the UK \cite{Case3,useche2026optimizing}.

The stability of multi-infeed, offshore AC islands depends heavily on the coordinated operation of multiple HVDC links and large offshore wind farms. Severe AC faults, DC faults, or HVDC link disconnections create large power imbalances, jeopardising the transient stability of the AC island along with potential consequences for the wider interconnected grids.
This fact underscores the necessity for robust fault management strategies that minimise the impact of malfunctions of the island, while at the same time meeting the security and quality of supply standards of national transmission systems~\cite{conf_paper}.

Several studies have explored Fault Ride Through (FRT) and stability related issues in HVDC and offshore wind dominated systems~\cite{mugambi2025methodologies}. Proposed methods include the use of braking resistors \cite{WF_FRT1}, excess power absorption via converter sub-module capacitances~\cite{WF_FRT2}, coordinated MMC–WTG control schemes~\cite{WF_FRT3}, offshore MMC voltage drop control for rapid active power reduction~\cite{WF_FRT4}, and other advanced control based approaches~\cite{WF_FRT5, WF_FRT6, WF_FRT7,raza2024fault}. However, these solutions predominantly address faults on onshore AC grids and are not adequate for the unique challenges faced by offshore AC islands with multiple HVDC infeeds.

In the literature, to the extent of the authors' knowledge, the topic of dynamic security assessment and control coordination for multi-infeed AC island systems remains largely unexplored.
 In~\cite{island_FRT1}, the authors investigate droop based power sharing strategies for multiple HVDC links following single link disconnections, but without time domain analysis of the transient impacts or system fault behaviour.
 In~\cite{island_FRT2}, a decentralised droop based emergency frequency control is proposed for LCC HVDC systems, but the focus is on AC side frequency events rather than fault induced transients.
 From the above, it can be concluded that research specifically targeting offshore energy islands, with emphasis on the time domain transient analysis and including both AC and DC faults occurring on the energy island remains limited.

This paper expands and elaborately validates the concept initially presented in~\cite{conf_paper}, addressing the presented research gap through a comprehensive fault management and stability enhancement coordination strategy for offshore AC islands with multiple HVDC infeeds. The proposed control approach coordinates HVDC links and WFs to manage both AC and DC faults effectively. 
In summary, the contributions of this paper are the following:

\begin{enumerate}
  \item A detailed transient stability assessment of multi-infeed offshore AC islands under both AC and DC fault conditions (including complete HVDC link disconnections), scenarios that have received limited attention in the existing literature.
  \item A coordinated, active power droop–based control strategy that jointly regulates the HVDC links and offshore WFs to manage severe disturbances, mitigating power imbalances and ensuring system resilience and stable operation during and after faults.
  \item A differentiated control framework for Grid-Following (GFL) and Grid-Forming (GFM) converters, ensuring that they operate within their current limits during the contingencies.
  \item Extensive EMT-level validation using PSCAD/EMTDC, together with a real-time laboratory setup, demonstrating the effectiveness of the proposed strategy in enhancing system robustness and ensuring stable post-fault performance.
\end{enumerate}

The paper is organized as follows: In Section II, the offshore AC island topology with HVDC in-feeds is presented while the proposed fault-ride-through coordination strategy and control design are discussed in Section III. Verification through extensive simulation results is provided in Section IV, while power hardware-in-the-loop (PHIL) experimental validation is evidenced in Section V. Section VI concludes the paper.

\begin{figure}[!t]
\centering
\includegraphics[width=1\columnwidth]{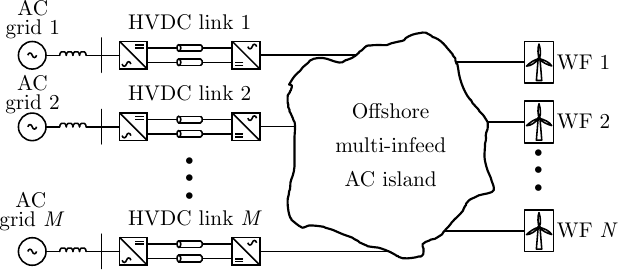}
%\vspace{-0.1cm}
\caption{Conceptual schematic of an offshore multi-infeed AC island with $M$ connected HVDC links and $N$ WFs.}
\vspace{-0.2cm}
\label{fig.offshore_island}
\end{figure}

\section{Offshore AC islands with HVDC in-feeds}

The core focus of this work is the transient stability analysis and fault operation of multi-infeed offshore AC energy islands, interconnected to the main grid via multiple HVDC links. A representative generic configuration is illustrated in Fig. \ref{fig.offshore_island}. These AC energy islands are predominantly inverter-dominated systems, comprising offshore WFs and HVDC converters operating under a combination of GFL and GFM control modes.
For the control mode assignment to the different participating converters, WFs are typically operated in GFL mode, while the necessary for maintaining stable voltage in the AC island GFM operation is provided by the HVDC links.
While sophisticated role-assignment methodologies can be found in the literature, in practical scenarios a single HVDC link with the highest rated capacity is tasked with operating in GFM mode, while the rest operate in GFL mode, which is the scenario considered in this work~\cite{soler2023interconnecting,arevalo2024scheduling}.
Finally, it should be noted that the HVDC links found in the AC energy islands can be of arbitrary configuration, including both symmetric and asymmetric monopole, as well as bi-pole configurations~\cite{van2016hvdc,puricelli2025bipolar}.
The variety of control modes, including heterogeneous FRT strategies, found in the converters of the AC island
may trigger complex interactions, potentially leading to adverse dynamic responses and instability during faults.
In addition, the impact of DC faults within the HVDC export infrastructure presents a critical and non-sufficiently studied challenge. Owing to the tight coupling between the HVDC links and the offshore AC network, disturbances originating on the DC side can propagate rapidly across the entire island, jeopardising system survivability.
The above highlight the necessity of a coordinated FRT response for all participating devices that is able to secure the safe recovery of the system after both AC and DC faults.

A particularly critical scenario arises following a fault on an HVDC link operating in export mode, where the affected converter is rapidly disconnected. This action produces an abrupt and substantial active power surplus within the offshore island, imposing severe stress on the remaining resources.  The challenge is further exacerbated when the GFM units, which are responsible for balancing power and stabilising frequency, are already operating close to their capability limits.
Furthermore, maintaining stability becomes increasingly difficult in configurations where the aggregate GFM capacity is comparable to that of the GFL-controlled units, leaving limited headroom for effective disturbance rejection.

Addressing these challenges necessitates advanced fault management and recovery strategies, capable of coordinating all the participating devices towards responding rapidly and adapting to a wide range of severe operating conditions, including undesirable power imbalances.
In this context, the present work develops and validates a communication-less, coordinated control and recovery strategy aimed at enhancing the stability and resilience of offshore AC islands subject to both AC and DC fault events.

\begin{figure}[!t]
\centering
\includegraphics[width=0.8\columnwidth]{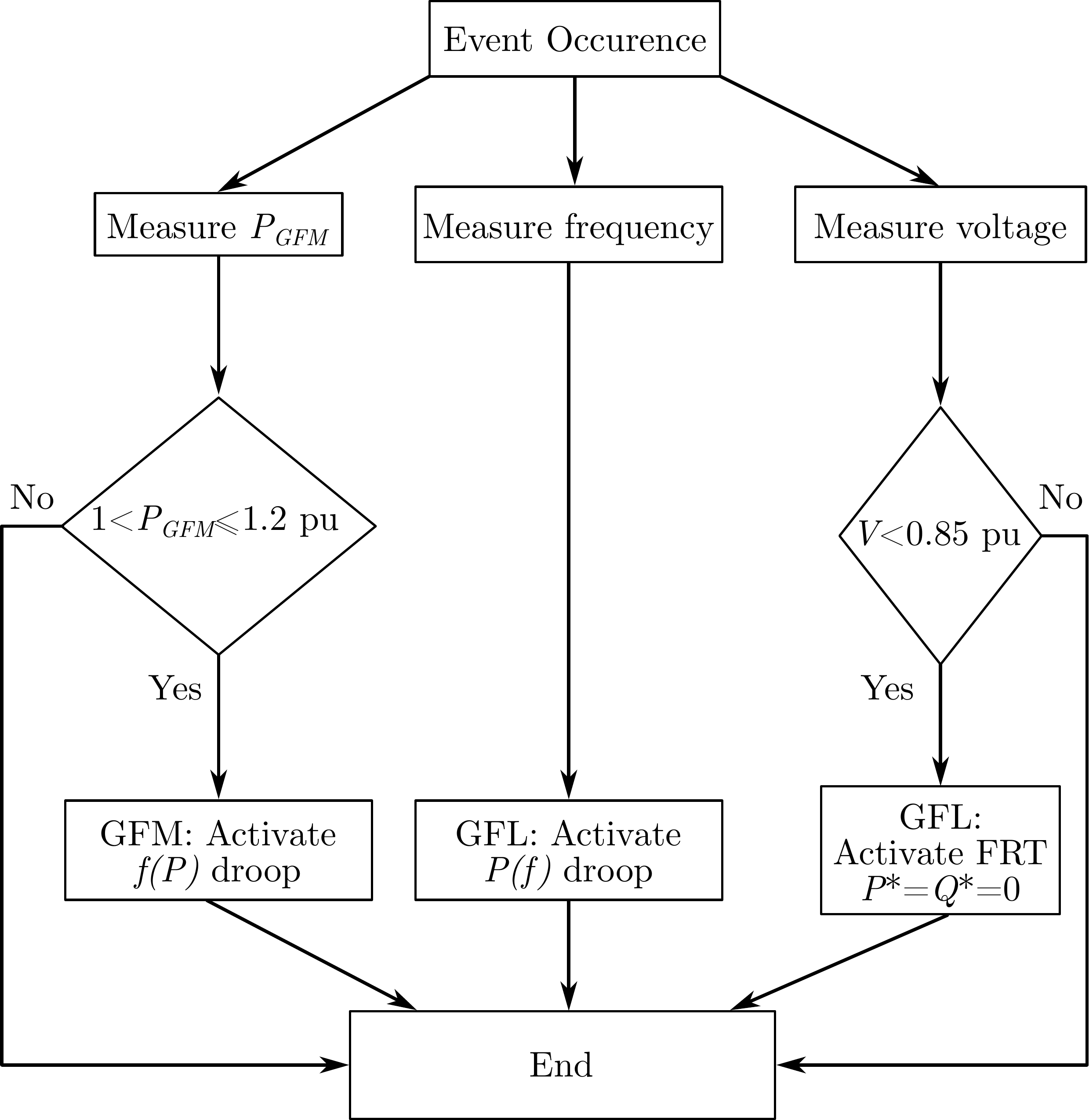}
%\vspace{-0.1cm}
\caption{Flowchart of the coordinated fault management strategy.}
%\vspace{-0.3cm}
\label{fig.flowchart}
\end{figure}
\begin{figure}[!t]
\centering
\includegraphics[width=0.8\columnwidth]{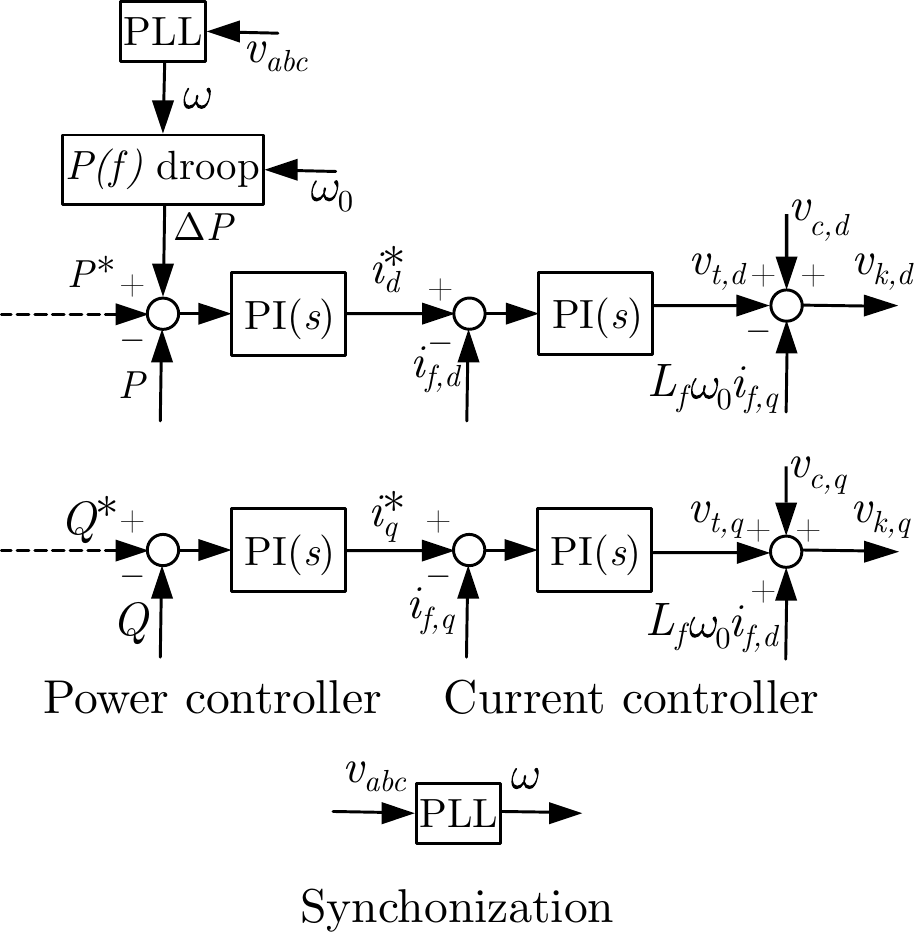}
%\vspace{-0.1cm}
\caption{Block diagram of GFL control structure. Dashed lines indicate parts of the control that are altered (activated/deactivated) during the FRT coordination algorithm.}
%\vspace{-0.3cm}
\label{fig.gfl_control}
\end{figure}
\begin{figure}[!t]
\centering
\includegraphics[width=0.8\columnwidth]{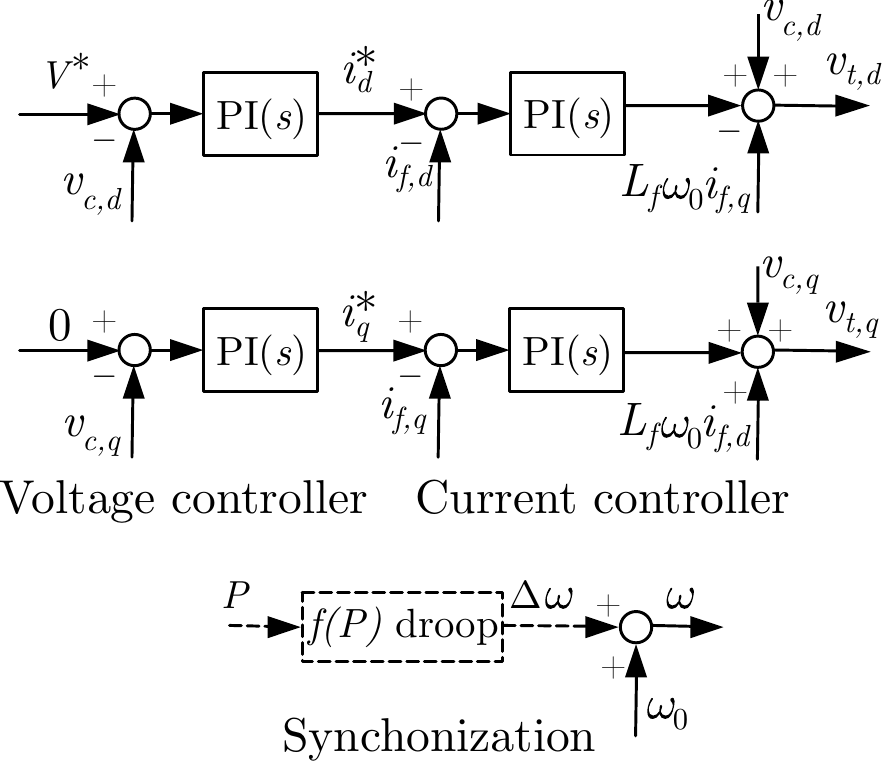}
%\vspace{-0.1cm}
\caption{Block diagram of GFM control structure. Dashed lines indicate parts of the control that are altered (activated/deactivated) during the FRT coordination algorithm.}
%\vspace{-0.3cm}
\label{fig.gfm_control}
\end{figure}

\section{Fault-Ride-Through Control Design}

The proposed FRT strategy for multi-infeed AC offshore islands incorporating HVDC links is presented in this section, which is the main contribution of this work.
%to ensure the stability and power balance .
%
This strategy particularly focuses on the response during and following the occurrence of different kinds of faults.
Specifically, both symmetric AC faults in the island, as well as DC faults in the GFL HVDC are considered.
For the former, the system is expected to return to its previous state after the fault is cleared, while for the latter, the DC fault is assumed to trip the whole HVDC link within the timescale of interest, thus leading to a new power flow within the AC island.
Throughout this work, it is considered that the HVDC link that operates in GFM mode remains active across all transient events.
It should be noted that unbalanced faults and negative sequence control lie outside the scope of this work.
In this regard, key adjustments are made to the generic GFL and GFM control structures,
%shown in Figs.~\ref{fig.gfl_control} and~\ref{fig.gfm_control},
as explained in the following subsections.
Finally, it should be noted that the proposed adjustments can be implemented to all types of HVDC configurations, without loss of generality.

\subsection{Response during fault}

The response during faults is a key aspect of the grid connection compliance FRT requirements in transmission systems~\cite{tsili2009review}.
The general requirement for onshore WF interconnection particularly focuses on the fast reactive current injection, within the limits of each converter, while maintaining a connection to the grid for a specified minimum amount of time.
Such an operation ensures that the system can return to its pre-fault condition smoothly, following to the event clearance.

In contrast, the proposed FRT strategy presented in this work involves suppressing both active and reactive power output of the GFL converters of both WFs and HVDC links during faults, as in:
\begin{equation} 
\label{eq:PQ_supression}
P^*=P^*_{FRT}, Q^*=Q^*=Q^*_{FRT}, \; \text{when} \; V<0.85 \; \text{pu},
\end{equation}
where $P^*$, $Q^*$ are the active and reactive power setpoints, respectively, of all GFL converters connected to the AC island, $V$ is the positive sequence voltage magnitude measured at their interconnection terminals and $P^*_{FRT}$, $Q^*_{FRT}$ are the updated active and reactive power setpoints during fault operation.

For the FRT coordination strategy proposed in this work, the fault setpoints are set as: $ P^*_{FRT}=Q^*_{FRT}=0$.
This power output suppression serves to mitigate the transient disturbances by limiting the current and voltage stresses on the AC offshore island, caused by the power imbalance during faults.
The proposed setup is also compatible with generic FRT setpoints that require reactive power injections during faults, which can be achieved simply by setting: $Q^*_{FRT} \neq 0$.

For the GFL HVDC links, the suppression of the active injection can be achieved in a very short timescale without implications for the internal power balance of the converter.
For the WFs, the suppression can be achieved by either adjusting the pitch angle of the wind turbines by means of the rotor speed control or by simply dissipating the excess active power~\cite{gao2026coordinated}.
The non-instantaneous response of these operations can be included in the study by properly adjusting the bandwidth of the outer active power loop of the GFL WFs, shown in Fig.~\ref{fig.gfl_control}, hence also motivating the choice to suppress the power references during the fault, instead of completely by-passing the outer control loops.
\subsection{Post-fault response}

For the post-fault response, two coordinated droop control strategies for both GFM and GFL converters are proposed.
These controllers aim to re-balance the power flow within the AC island after the contingency, taking into account the remaining power injections from the WFs and the HVDC links.

For the GFM converter of the HVDC link, an $f(P)$ droop control is proposed that adjusts the offshore island frequency based on the power imbalance post-fault.
This $f(P)$ droop control is defined as:
\begin{equation} 
\label{eq:GFM_droop}
\omega_{GFM} = \omega_{0} - K_{GFM} \times P_{GFM}, 
\end{equation}
where $P_{GFM}$ is the active power output of the GFM converter, $\omega_{0}$ indicates the nominal angular frequency and $K_{GFM}$ is the droop gain of the GFM $f(P)$ droop control.  

The frequency update caused by the GFM droop controller is used as a triggering signal to the GFL converters of the AC island, which curtail their power during the post-fault recovery period, as required to re-balance the power flow within the network.
This curtailment is implemented by means of corresponding outer $P(f)$ droop controllers, which measure the adapted frequency of the network and respectively adapt their own power output according to:
\begin{equation} \label{eq:GFL_droop}
\Delta P = K_{GFL} \times (\omega_{0} - \omega_{PLL}),
\end{equation}
where $\Delta P$ is the corrective reference signals for the active power controller of the GFL converters,  $K_{GFL}$ is the $P(f)$ droop gain, and $\omega_{PLL}$ is the angular frequency measured through the PLL device of the GFL converters.
 %, all expressed in rad/sec.
%
As it can be observed, by means of the frequency measurement performed by the PLL,
%upon effective PLL operation of the GFL control system,
%
$\omega_{PLL}$ is equal to $\omega_{GFM}$, therefore highlighting the principles of the proposed signaling strategy, i.e., the use of the offshore AC island frequency as a means of fault response coordination without requiring a dedicated communication link.
Additionally, the corresponding power re-balance by the offshore GFL converters, based on the updated frequency signal, further assists system recovery in scenarios that may cause imbalances in active power distribution, such as the disconnection of an HVDC link. 

\subsection{Droop gain calculation}

The proposed strategy considers a static adjustment the $K_{GFL}$ gain of the $P(f)$ droop controller of the GFL WF converters, according to each active power dispatch in the network. This adjustment is crucial for maintaining stability during the worst case event considered in this work (i.e., the disconnection of all the GFL HVDC links in exporting mode), and is assumed to be supported by a higher-level steady-state control scheme. Even though this scheme is not explicitly modelled in the experimental validation of the following sections, the rules for calculating the droop gains at each steady-state condition are provided in this subsection.

Assuming that $P_{exp}$ is the total active power exported by the HVDC links operating under GFL control, under a complete disconnection of all these links, the power output of the WFs should be curtailed to preserve the power balance within the AC island. Assuming there are $N$ WFs integrated within the AC offshore island, and that the GFM inverter can be overloaded by up to an additional 0.2 p.u. of its nominal capacity, the power balance after the fault clearance is described by:
\begin{equation} \label{eq:power_balance}
N\Delta P= P_{exp}-(1.2P_{GFM}^N-P_{GFM}^0),
\end{equation}
where $P_{GFM}^N$ is the nominal capacity of the GFM inverter,  and $P_{GFM}^0$ is the pre-fault operation point. Substituting~\eqref{eq:GFM_droop}-\eqref{eq:GFL_droop} into \eqref{eq:power_balance} leads to the analytical calculation of the GFL droop gain as:  
\begin{equation} \label{eq:GFL_droop_gain}
K_{GFL}=\frac{P_{exp}-(1.2P_{GFM}^N-P_{GFM}^0)}{NK_{GFM}1.2P_{GFM}^N}.
\end{equation}
In order to differentiate normal operating conditions from transient events, impacting the power balance of the power island, the $P(f)$ droop of the GFM converter is only activated within the deadband defined by the nominal capacity and the thermal limit of the GFM (1 pu and 1.2 pu, respectively, for most of the typical HVDC converters), as shown in Fig.~\ref{fig.flowchart}. Therefore, GFM output power within this range serves as a critical indicator of either asset loss or an imbalance in the power distribution on the offshore island, triggering the necessary droop control-based curtailment functionality to address these issues dynamically.

Under this droop design, for every $N\Delta P>0$, the GFM will operate at it's current limit, i.e., at 1.2 p.u., triggering a frequency increase (e.g., to 51.5 Hz under a $K_{GFM}=3\%$), leading to the WFs appropriately curtailing their power output to ensure balance within the offshore network. 
If the GFM does not reach its thermal limit but still enters an overloaded operation, the frequency will still increase, leading to the WFs curtailing their active power according to the equilibrium point resulting from~\eqref{eq:GFM_droop} and~\eqref{eq:GFL_droop}.
The above sequence of events ensures the resilience of the system, until the disconnected link is restored, or new references are received through the higher level steady-state controller. 

Fig.~\ref{fig.flowchart} illustratively summarizes the comprehensive coordination FRT strategy presented in this work, while Figs.~\ref{fig.gfl_control} and~\ref{fig.gfm_control} present the control schemes used in this work for the GFL and GFM converters, respectively. It can be seen that they consist of standard cascaded current and outer control loops (power and voltage loops for the GFL and the GFM, respectively), enhanced by the FRT adaptations presented in this work~\cite{pogaku2007modeling,collados2019stability}.

\section{Simulation verification}
\label{sec.simulation}

\begin{figure*}[!t]
\centering
%\vspace{-0.1cm}
\includegraphics[width=0.85\textwidth]{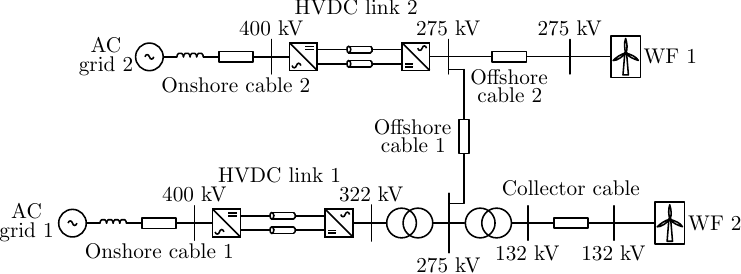}
%\vspace{-0.4cm}
\caption{Diagram of the multi-infeed network under study.}
%\vspace{-0.3cm}
\label{fig.system_overview}
\end{figure*}

In this section, the proposed control strategy is validated in a PSCAD/EMTDC simulation environment against different fault conditions.
The simulated network is illustrated in Fig.~\ref{fig.system_overview}. As shown, the system comprises two HVDC links, each rated at 1200MW. HVDC‑1 operates in GFM control mode, using the controller showcased in Fig.~\ref{fig.gfm_control}, whereas HVDC‑2 operates in GFL mode, using the controller showcased in Fig.~\ref{fig.gfl_control}. On the onshore side, the terminals of the two HVDC links are connected to separate 400kV interconnection points, while their offshore terminals are connected to a common offshore AC island at 275kV.
The offshore system also includes two WFs connected to the island network at 275kV (WF1) and 132kV (WF2), respectively, with capacity of 1200MW each. Both WFs operate in GFL mode. A detailed description of the modelling approach and the electrical and control specifications of the Modular Multilevel Converters (MMCs) used for the HVDC links and the WF converters can be found in~\cite{psaras2020non}.

\subsection{AC fault}
The first scenario investigates a symmetric three‑phase AC fault applied to the 275~kV bus of the AC offshore island, located at the low-voltage side of the transformer of HVDC link~1. The fault is initiated at t = 10.4~s, lasts for 300~ms, and does not trigger the disconnection of any generating unit. Prior to the disturbance, all units operate at their maximum output: GFM HVDC‑1 exports 1200~MW to the onshore grid, GFL HVDC‑2 exports 1200~MW, and both WFs deliver 1200~MW collectively.

\begin{figure}[!t]
\centering
\includegraphics[width=1\columnwidth]{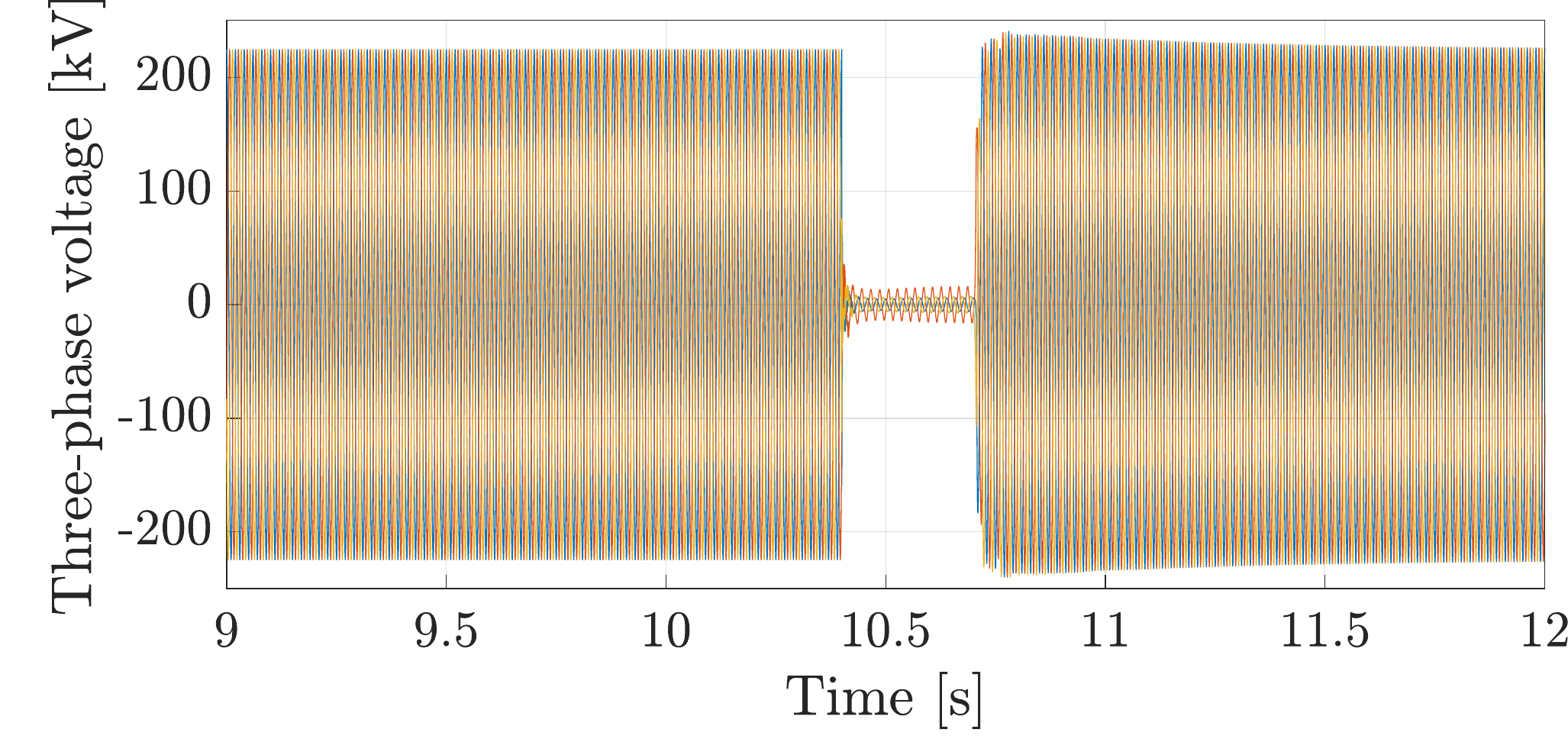}
\caption{Three-phase voltage during the AC fault.}
\vspace{-0.2cm}
\label{fig.simulation_ac_fault_voltage}
\end{figure}

\begin{figure}[!t]
\centering
\includegraphics[width=1\columnwidth]{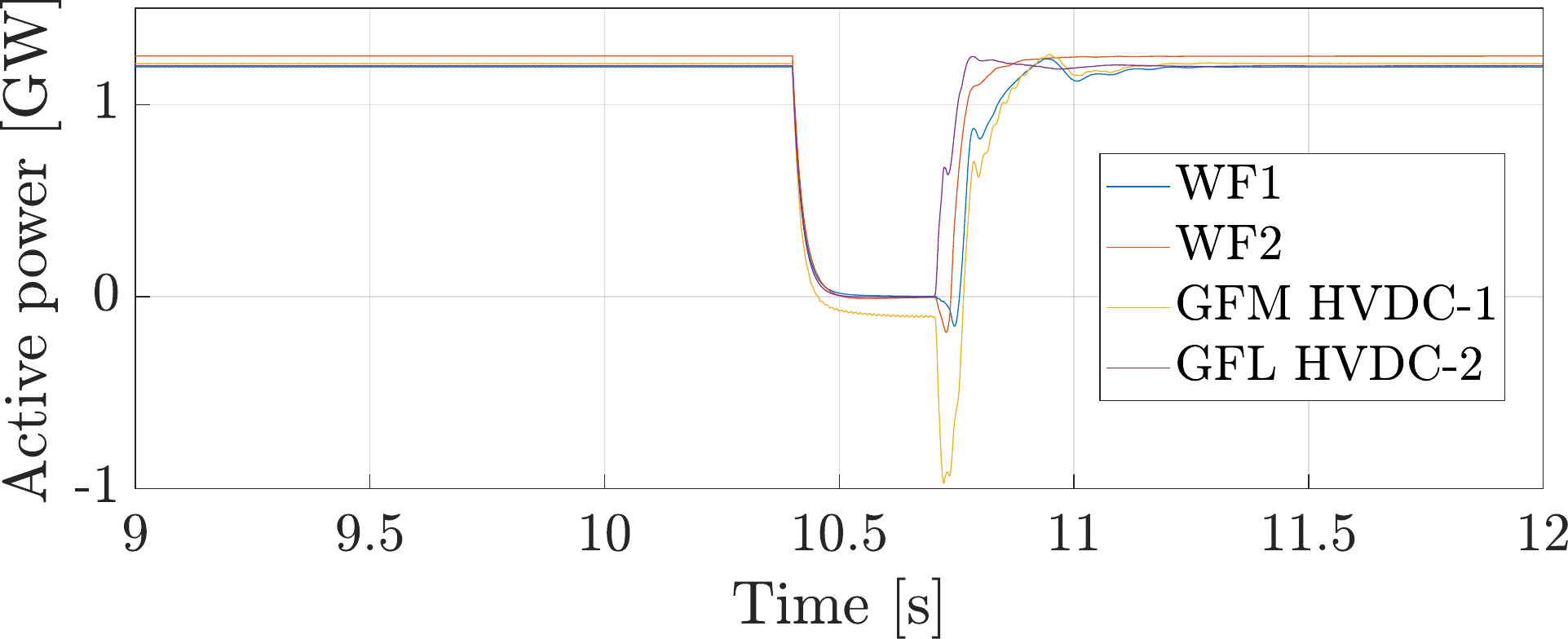}
\caption{Converter active power during the AC fault.}
\vspace{-0.2cm}
\label{fig.simulation_ac_fault_active_power}
\end{figure}

Fig.~\ref{fig.simulation_ac_fault_voltage} shows the three-phase voltage of the 275~kV bus where the fault is applied.
It can be seen that the voltage value falls to approximately to 0~p.u. at the fault onset, while, once the fault is cleared, it rapidly recovers.
%, and all units return quickly to their pre‑fault power levels.
During the fault, HVDC‑2 and the WFs demonstrate their FRT capability, shown in Fig.~\ref{fig.simulation_ac_fault_active_power}, with all the GFL converter active power injection dropping  to zero during the fault, followed by a rapid recovery to the pre‑fault power levels after the fault clearance.

This scenario highlights the effectiveness of the implemented FRT control strategy. Limiting the current contribution during the disturbance enables a fast post‑fault recovery and enhances overall system stability. Furthermore, the results demonstrate the selective behaviour of the control approach: it remains passive during AC faults that do not affect the active power dispatch thereby preserving operational robustness.


\subsection{DC fault}
\begin{figure}[!t]
\centering
\includegraphics[width=1\columnwidth]{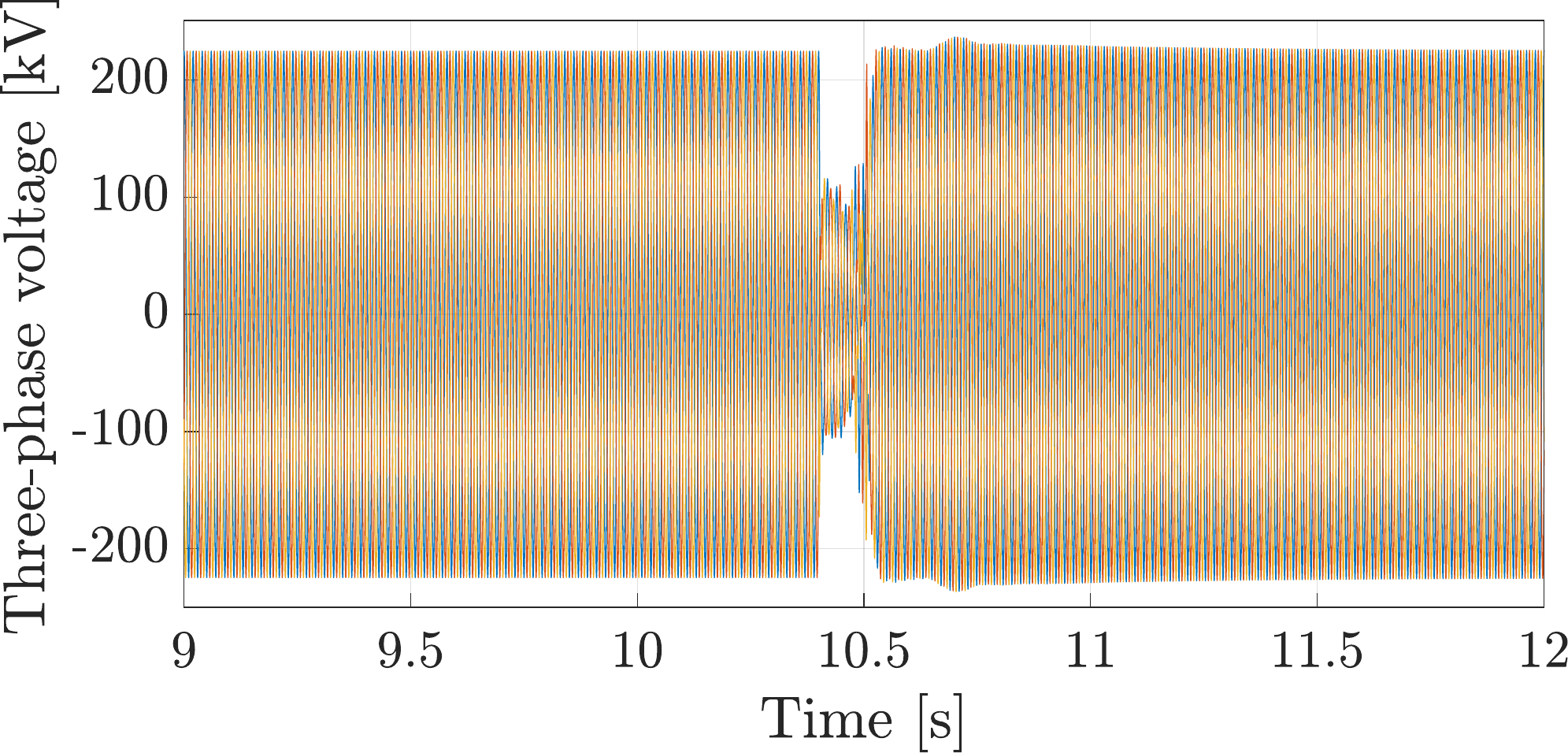}
\caption{Three-phase voltage during the DC fault.}
\vspace{-0.2cm}
\label{fig.simulation_dc_fault_voltage}
\end{figure}

The second scenario examines a pole‑to‑pole DC fault occurring on the GFL HVDC link (HVDC‑2) at t = 10.4~s, which results in the disconnection of HVDC‑2 within approximately 250~ms.
The DC fault results in a voltage dip at the 275~kV bus, located at the low-voltage side of the transformer of HVDC link~1, which is shown in Fig.~\ref{fig.simulation_dc_fault_voltage}.
Prior to the fault, HVDC‑2 was operating at its maximum export capability of 1200~MW, transferring power from the offshore AC island to the onshore grid (represented by a negative power sign). Both offshore WFs (WF1 and WF2) were also producing their rated active power, namely 1200~MW each. The remaining 1200~MW required to maintain power balance was exported through HVDC‑1, which was operating in GFM mode and at its rated capacity. Thus, all generation units, including the GFM converter, were operating at or near their limits before the disturbance occurred.

\begin{figure}[!t]
\centering
\includegraphics[width=1\columnwidth]{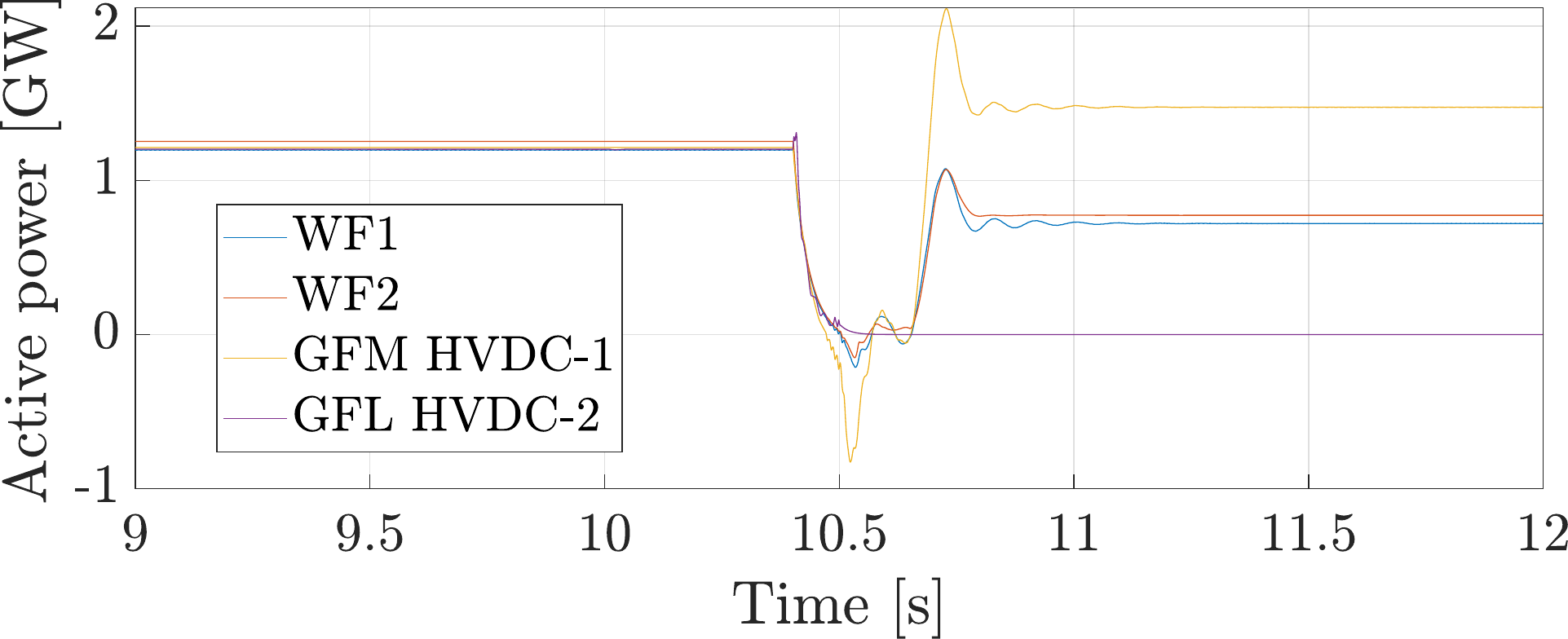}
\caption{Converter active power during the DC fault.}
\vspace{-0.2cm}
\label{fig.simulation_dc_fault_active_power}
\end{figure}

\begin{figure}[!t]
\centering
\includegraphics[width=1\columnwidth]{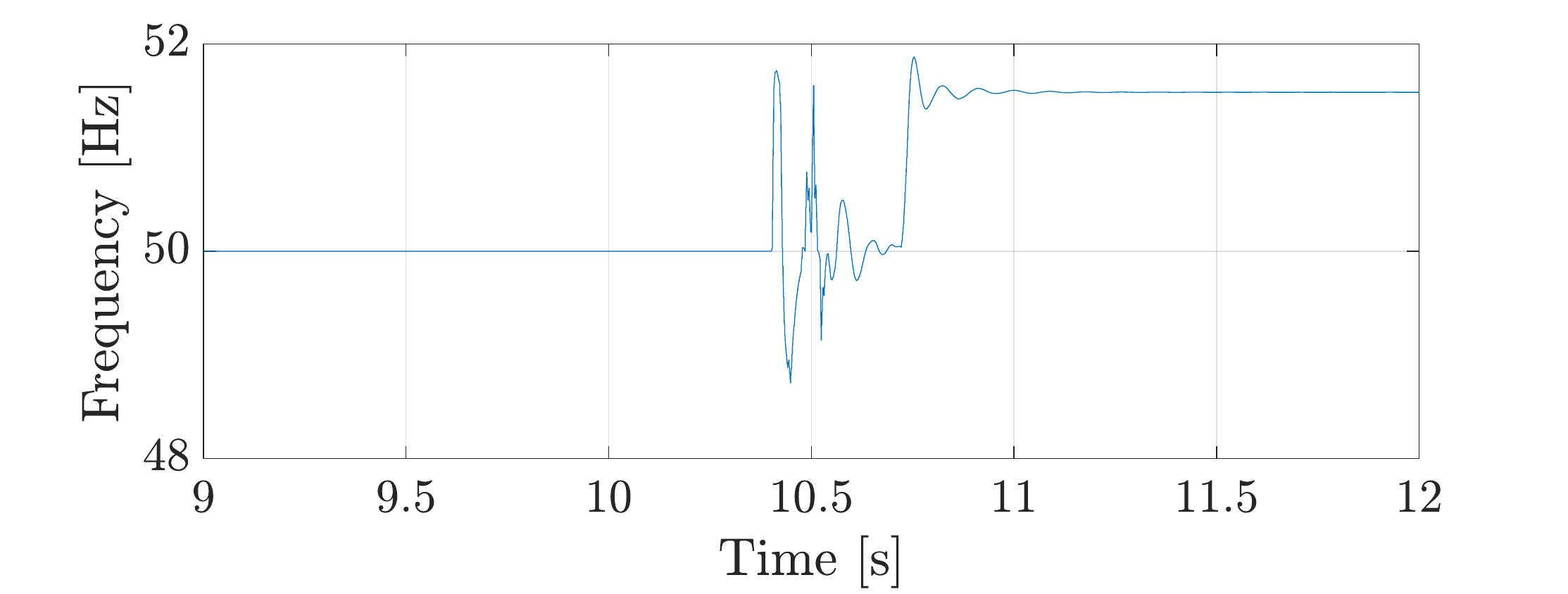}
\caption{AC island frequency during the DC fault.}
\vspace{-0.2cm}
\label{fig.simulation_dc_fault_frequency}
\end{figure}

Immediately following the DC fault, both WFs entered FRT mode, rapidly reducing their output to zero to alleviate transient stress on the offshore AC system.
After the loss of GFL HVDC‑2, GFM HVDC‑1 starts to operate in an overloaded condition, reaching approximately 1.2~pu (approximately 1440~MW export). Consequently, the outputs of WF1 and WF2 were curtailed symmetrically to approximately 720~MW each, consistent with their identical power ratings and same droop gains.
This operation both during and after the fault is illustrated in Fig.~\ref{fig.simulation_dc_fault_active_power}, where the active power injections of all participating converters is depicted. 
Within roughly 250~ms, HVDC‑2 was fully disconnected, leaving the system with a sudden 1200~MW surplus, originating from the WFs generation.
This surplus activated the frequency droop response of the GFM converter, resulting in an increase in system frequency to approximately 51.5Hz as a direct result of the GFM overloading and subsequent engagement of the corresponding $P(f)$ droop mechanisms in the WFs that re-balance the active power flow.
The frequency signal of the AC island before, during and after the fault is shown in Fig.~\ref{fig.simulation_dc_fault_frequency}, which illustrates the operation explained above.
Overall, these results confirm that the proposed strategy provides a rapid, coordinated, and stable response during a severe disturbance.


\section{Experimental Validation}

\begin{figure}[t]  % * makes it span across two columns, [b] places it at the bottom of the page
\centering
\includegraphics[ height=6cm]{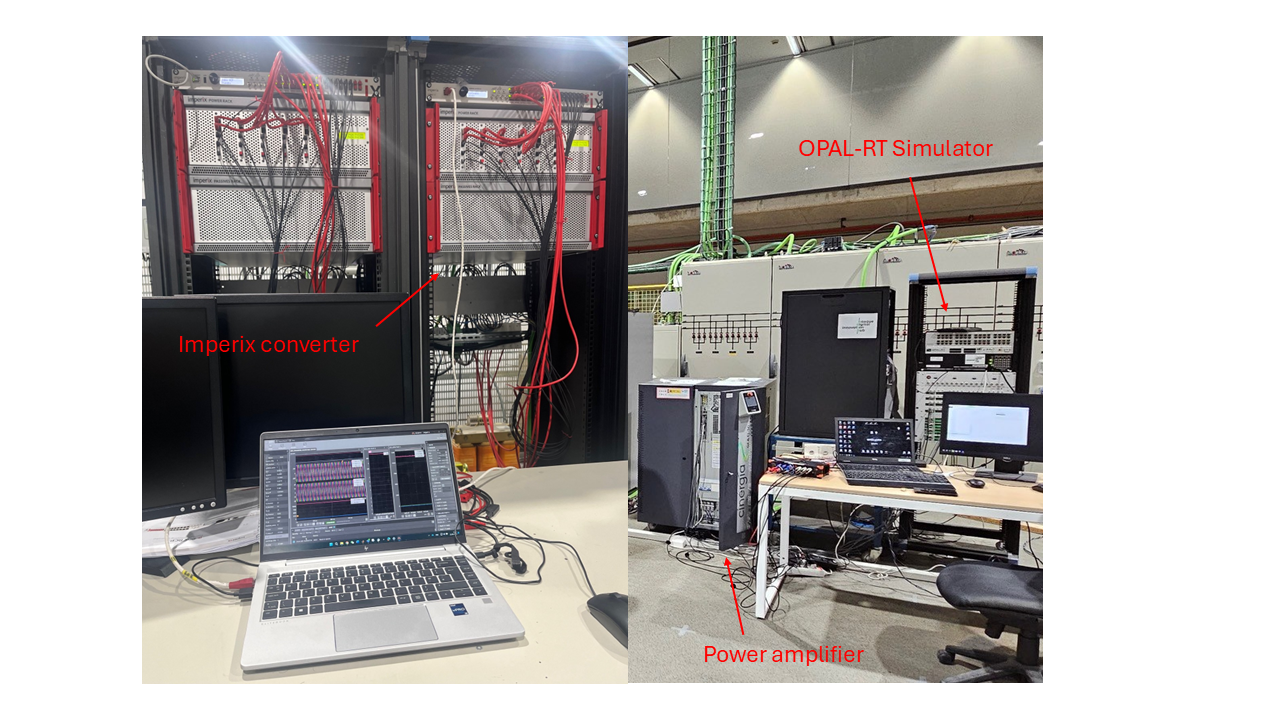}  % Adjust height as needed
\caption{PHIL laboratory setup.}
\label{fig:lab_setup}
\end{figure}

\subsection{Laboratory Setup Description}
In this section, a Power-Hardware-In-Loop (PHIL) simulation setup is utilized to validate the proposed control framework.
This section aims to particularly assess the performance of the developed strategy under real-world conditions, using actual hardware, namely real converters and their respective controllers.
Similar to the simulation studies, presented in Section~\ref{sec.simulation}, the experimental validation investigates two representative disturbance events, an AC fault at the offshore island and a DC fault at one of the HVDC links.
The experimental setup, designed to progressively validate the coordinated control strategy for AC offshore islands with multi-indeed HVDC and offshore wind farms, under different fault conditions and hardware configurations, is illustrated in Fig. \ref{fig:lab_setup}.

For the use case of this project, two HVDC links operating in both GFM and GFL modes, denoted as GFM HVDC-1 and GFL HVDC-2, respectively, are simulated within a MATLAB/Simulink environment, executed in real time on an OPAL-RT device, shown in Fig.~\ref{fig:lab_setup}.
Within the OPAT-RT model, an additional WF is also simulated (WF1).
A hardware Imperix operated in GFL mode with capacity of 15kW and rated voltage of 230V, is interfaced to the simulation model through a Cinergia power amplifier, emulating a wind-farm (WF2).
It is noteworthy that the high voltage network in the simulation environment is appropriately interfaced with the low-voltage hardware setup through an appropriate interface mechanism.
This setup allows for the dynamic interaction between the physical and simulated assets, even if the voltage and power ratings of the hardware converter are scaled down.
In more detail, both DC and AC faults can be realized through the simulated environment, and the response of a hardware GFL converter is tested, along with all the developed control algorithms, as presented in the following subsections.

\subsection{FRT Control Validation}

The results presented in this section showcase two examples of characteristic fault events that validate the efficacy of the proposed fault management algorithm.
In particular, the events include:

\begin{itemize}

\item 
A symmetric, three-phase AC fault applied at the common bus of the HVDC link, which does not lead to the disconnection of any unit.

\item 
A DC fault applied to one of the HVDC links, during which the offshore converter operates in GFL control mode and exports active power.
%in export mode.
This fault causes the disconnection of the HVDC link, leading to a readjustment of the power flow within the whole system.

\end{itemize}

It should be noted that when an HVDC link operates in export mode (indicated by a negative sign in active power), it supplies active power from the offshore AC island to the main grid. 
Conversely, in import mode (positive active power), the HVDC link absorbs active power from the main grid into the offshore AC island.

\subsubsection{AC Fault}
\begin{figure}[!t]
\centering
\includegraphics[width=1\columnwidth]{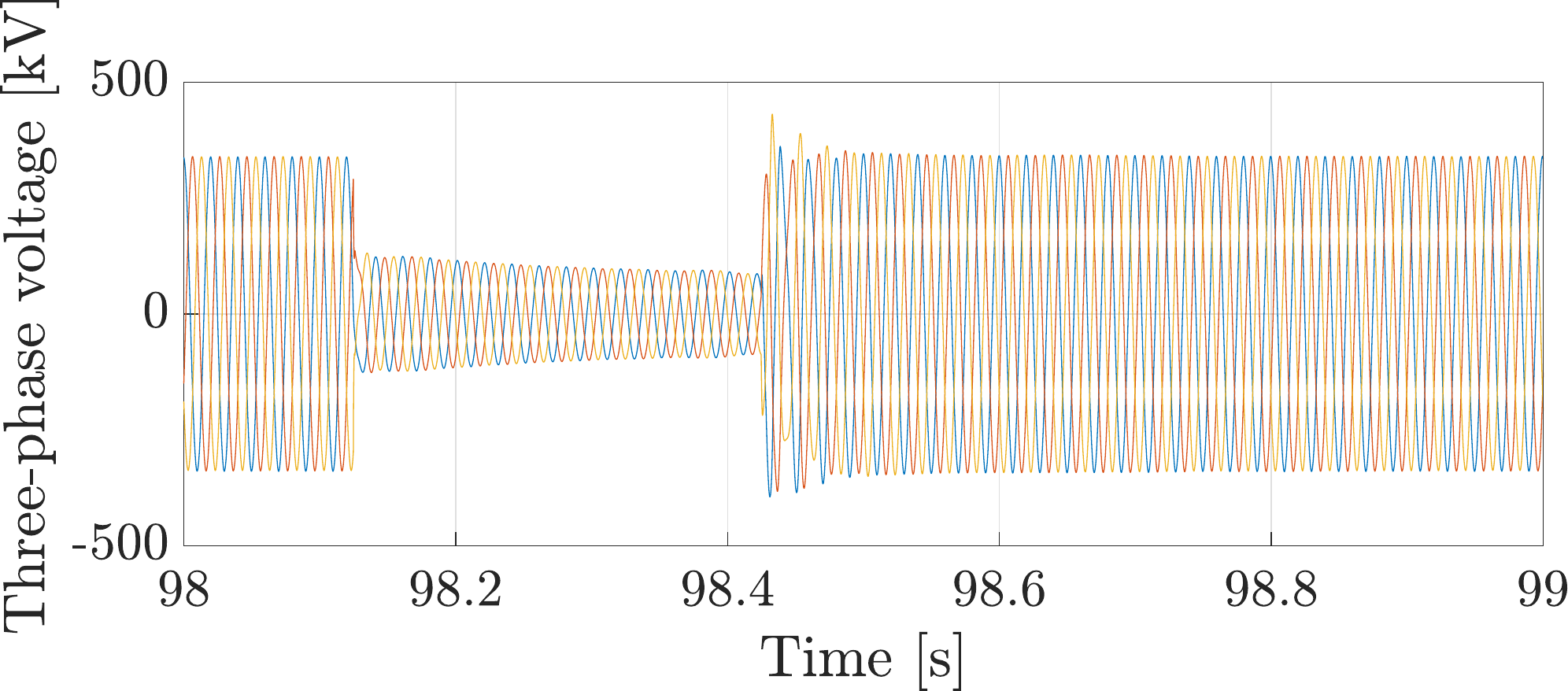}
\caption{Lab validation for AC fault. Three-phase voltage waveforms at the common connection point.}
\vspace{-0.2cm}
\label{fig.lab_ac_voltage}
\end{figure}
\begin{figure}[!t]
\centering
\includegraphics[width=1\columnwidth]{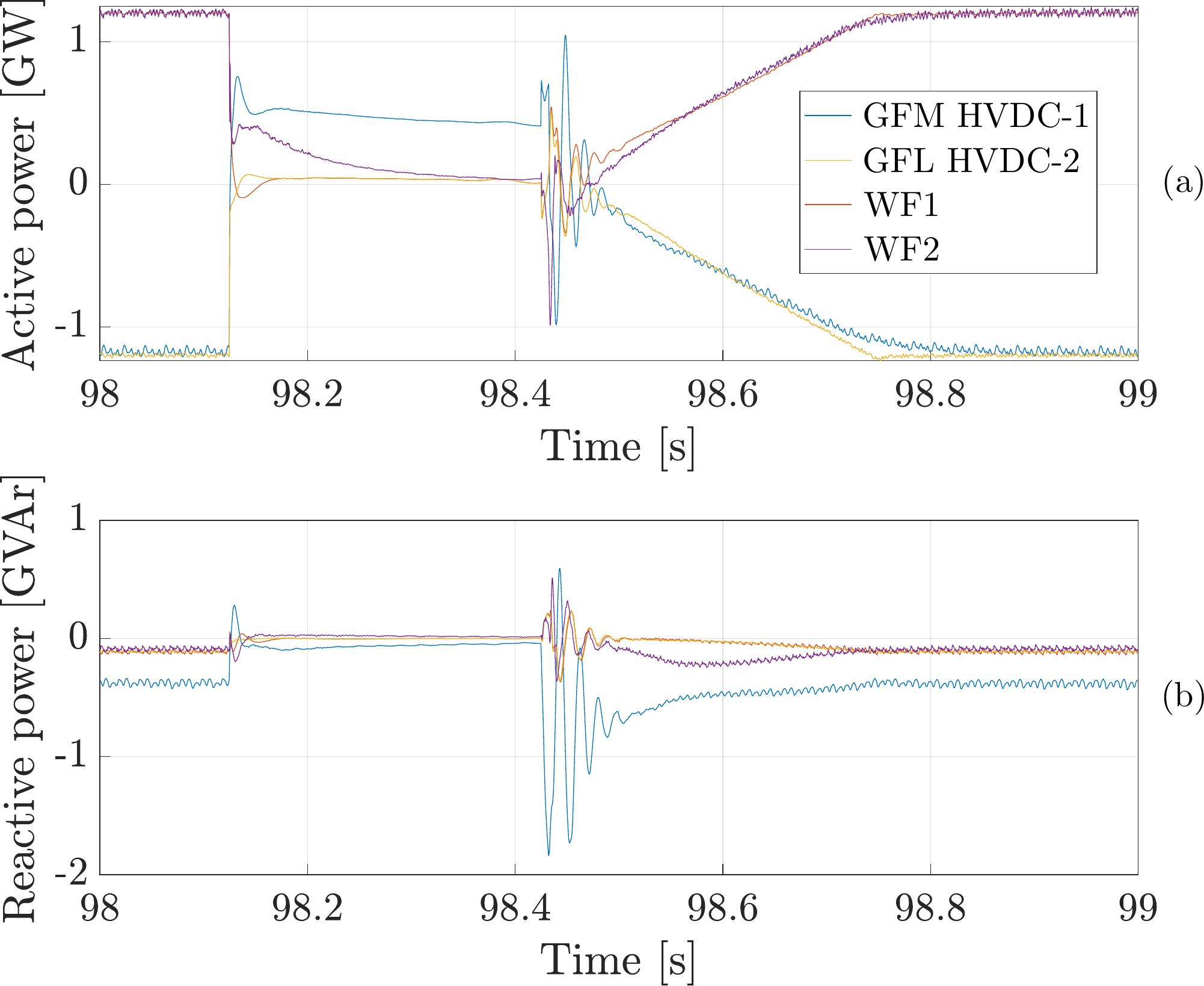}
\caption{Lab validation for AC fault. (a) Active and (b) reactive power injections from all participating converters.}
\vspace{-0.2cm}
\label{fig.lab_ac_power}
\end{figure}
\begin{figure}[!t]
\centering
\includegraphics[width=1\columnwidth]{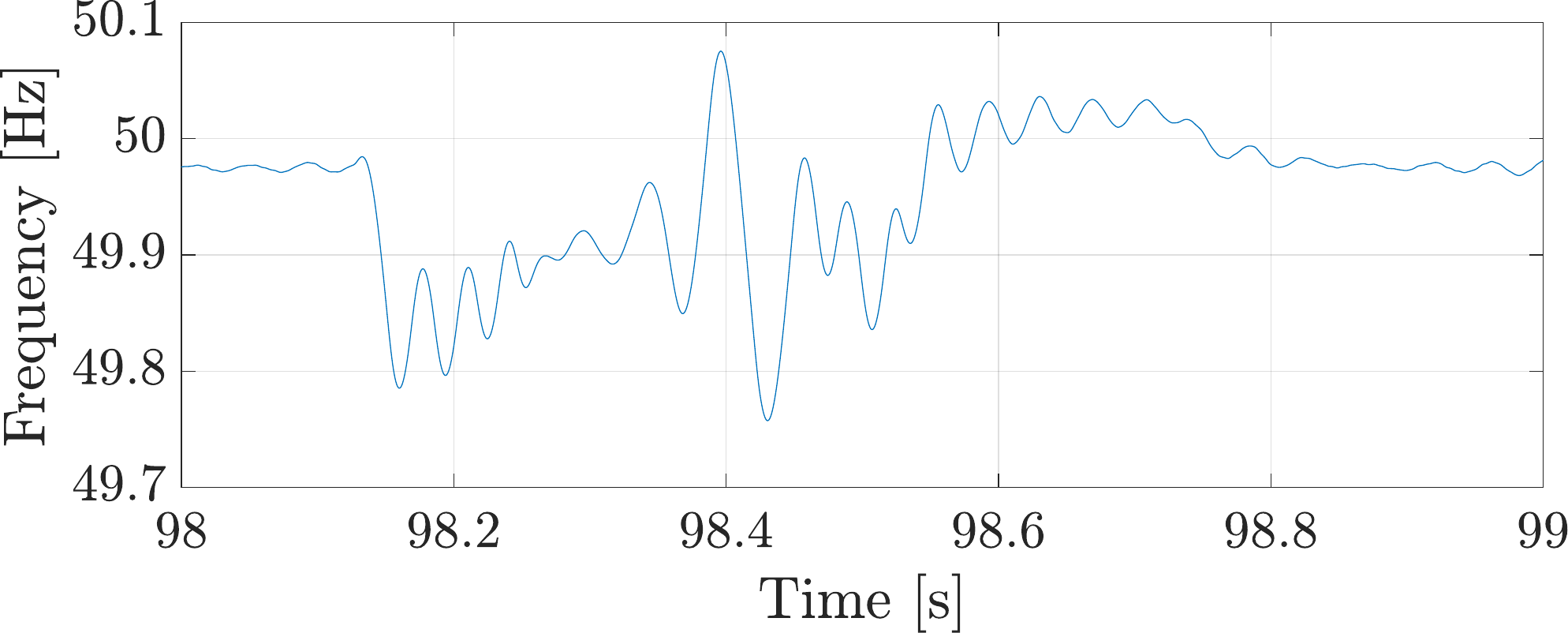}
\caption{Lab validation for AC fault. System frequency.}
\vspace{-0.2cm}
\label{fig.lab_ac_frequency}
\end{figure}

This scenario examines a solid three-phase-to-ground AC fault at the offshore AC island.

Prior to the fault, the system was operating at maximum capacity: WF1 and WF2 were each importing 1.2~GW, GFL HVDC-2 was exporting 1.2~GW, and an additional 1.2~GW was being exported by the HVDC link with the GFM converter, which operates as the power balancer.
The fault occurs at $t =$ 98.11~s and lasts for 300 ms.
Fig.~\ref{fig.lab_ac_voltage} shows the three-phase voltage waveforms at the common connection point of the HVDC links.
It can be seen that the applied fault causes a voltage drop to approximately 31\% of its original value. 

Fig.~\ref{fig.lab_ac_power} shows the active and reactive power injections from all participating converters.
It can be seen that all the GFL converters, equipped with the proposed FRT algorithm, quickly set their power setpoints to zero in order to not contribute further to the system stress.
Once the fault is cleared, all units promptly recover to their pre-fault power levels.
Finally, Fig.~\ref{fig.lab_ac_power} shows the frequency of the system during the event.
It can be seen that, despite the fluctuations during the fault, it is restored to its nominal value.
This implies that the droop control remains inactive, as designed for this kind of fault events, showcasing the selective nature of the proposed control strategy.
From the above, it can be concluded that the fault does not lead to the disconnection of any assets, allowing the active power dispatch of the system to remain unaltered during the post-fault operation.

\subsubsection{DC Fault}

This scenario examines the effectiveness of the developed algorithm during a pole-to-pole DC fault at the terminals of the offshore converter GFL HVDC-2, which results in the disconnection of the link within 250 ms.
Prior to the fault, the active power dispatch of the system was similar to the previous case.

\begin{figure}[!t]
\centering
\includegraphics[width=1\columnwidth]{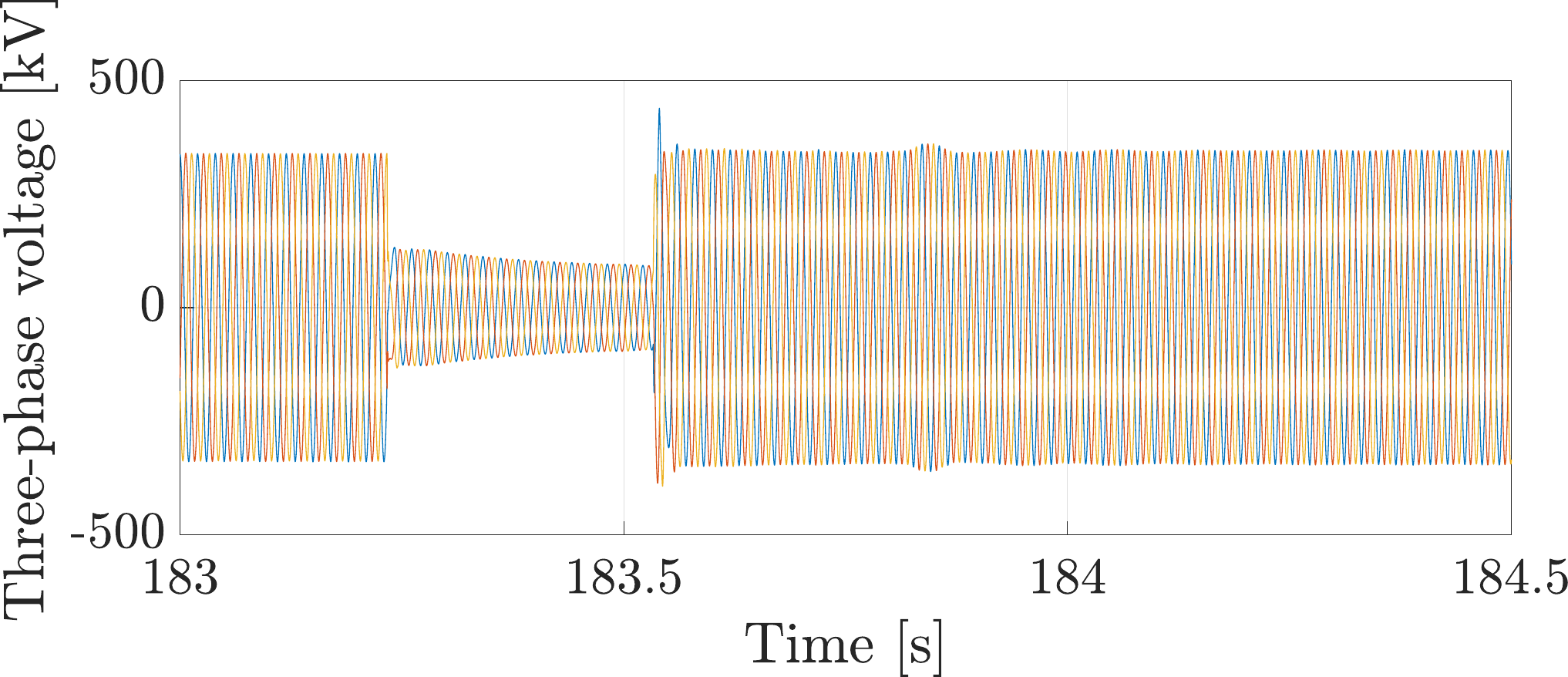}
\caption{Lab validation for DC fault. Three-phase voltage waveforms at the common connection point.}
\vspace{-0.2cm}
\label{fig.lab_dc_voltage}
\end{figure}

\begin{figure}[!t]
\centering
\includegraphics[width=1\columnwidth]{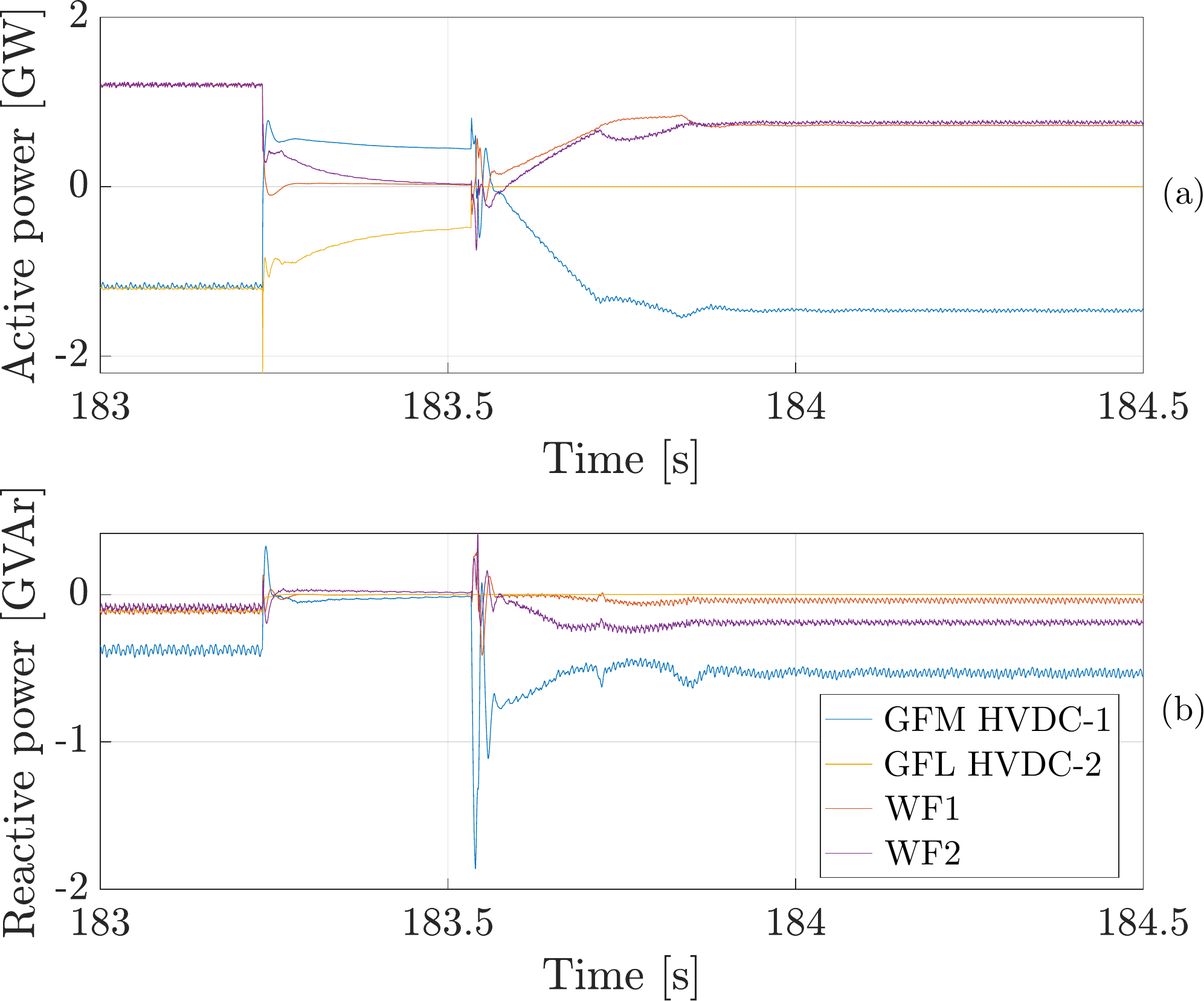}
\caption{Lab validation for DC fault. (a) Active and (b) reactive power injections from all participating converters.}
\vspace{-0.2cm}
\label{fig.lab_dc_power}
\end{figure}

The fault occurs at $t =$ 183.3~s, causing a voltage drop that is shown in Fig.~\ref{fig.lab_dc_voltage}.
Upon fault occurrence, WF1 and WF2 initiate FRT control, reducing their power output to zero to alleviate stress on the offshore AC island.
Within 250 ms, GFL HVDC-2 is disconnected, resulting in a power surplus of 1.2~GW, equivalent to the nominal capacity of the GFM unit.
The above are evidenced by Fig.~\ref{fig.lab_dc_power}, which showcases the active and reactive power injections from all system converters during the DC fault.

\begin{figure}[!t]
\centering
\includegraphics[width=1\columnwidth]{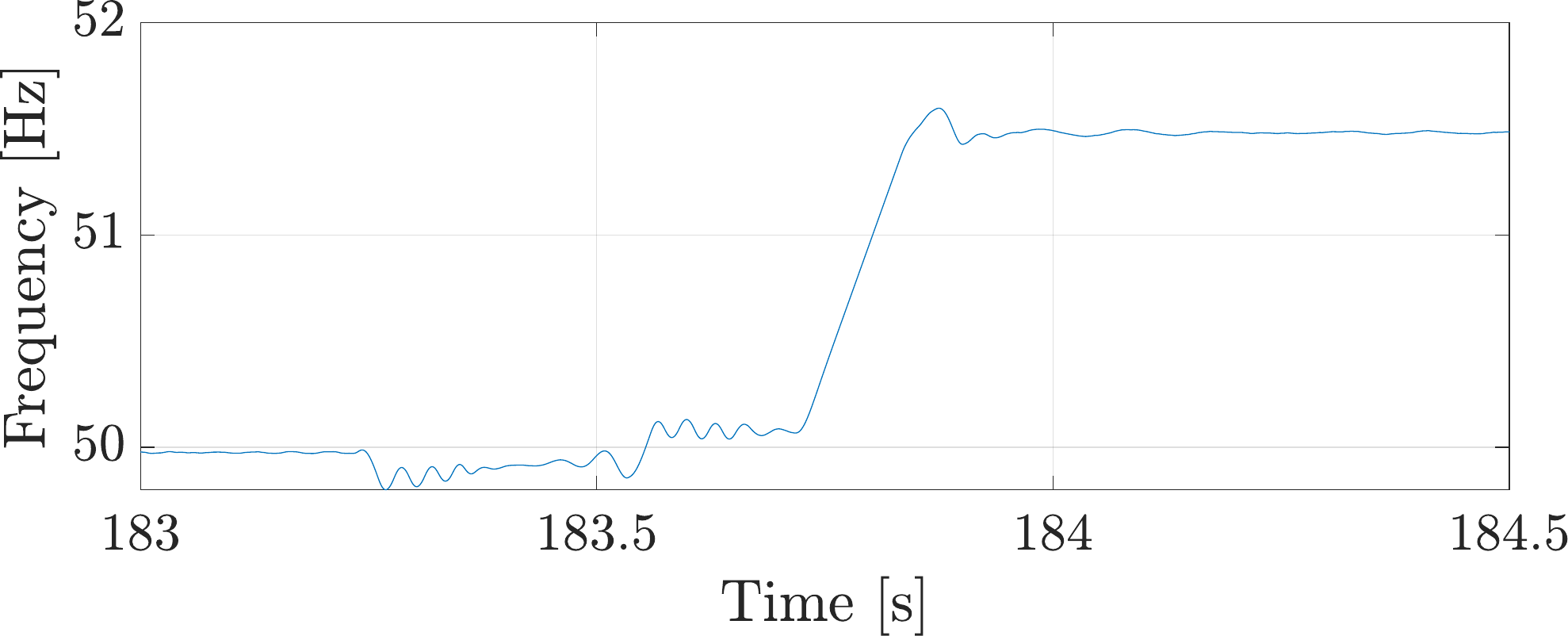}
\caption{Lab validation for DC fault. System frequency.}
\vspace{-0.2cm}
\label{fig.lab_dc_frequency}
\end{figure}

In close succession, this sudden surplus triggers the $f(P)$ droop response of the GFM unit, causing an increase in system frequency, shown in Fig.~\ref{fig.lab_dc_frequency}.
The frequency reaches 51.5 Hz due to the overload condition on the GFM unit, serving as a key indicator for the $P(f)$ droop response in WF1 and WF2.
In turn, this activates the corresponding $P(f)$ droop control in both WFs, whose effect can be inspected in Fig.~\ref{fig.lab_dc_power} after approximately $t=$ 183.5~s.
The activation of droop control in the GFM unit leads to its overloading, with its power export reaching 1.44~GW (exceeding its thermal limit of 1.2 pu).
As the frequency continues to rise, the $P(f)$ droop control in the WF converters is engaged, resulting in curtailment of their output to 0.72~GW each in order to stabilise the system and restore the power balance.

From the above, it can be concluded that the proposed algorithm successfully re-balances the power flow in the system, despite the loss of one generating unit.
This leads to the restoration of the AC voltage to its nominal levels after the isolation of the fault event.

\section{Conclusion}
This paper presented a coordinated fault-ride-through control strategy for multi-infeed offshore AC islands with HVDC interconnections, addressing critical challenges associated with fault operation and Grid Code compliance. By incorporating advanced FRT functionalities, including zero power injection during faults and coordinated post-fault droop control, the proposed approach effectively manages power imbalances, particularly under HVDC disconnection scenarios. Comprehensive validation through EMT simulations and PHIL experiments confirms the practicality and robustness of the proposed solution, supporting its suitability for future large-scale offshore wind farm integration.

\section*{Acknowledgments}

This work was supported by the European Union’s Horizon 2020 research and innovation programme, under Grant 870620, through using the ERIGrid 2.0 Research Infrastructure. It was also partially funded by CETPartnership, the Clean Energy Transition Partnership under the 2022 joint call for research proposals, co-funded by the European Commission under the R\&D project CET Wind Digipower (PCI2023-145953-2). The authors would like to particularly thank Asier Gil De Muro-Zabala, Angel Luis Perez-Basante and Igor Gabiola-Antxustegi from Tecnalia (Derio, Spain), for their excellent support during the experimental validation of the proposed algorithm.

\ifCLASSOPTIONcaptionsoff
  \newpage
\fi

\bibliography{refs}
\bibliographystyle{IEEEtran}

\end{document}